# Power-Distortion Metrics for Path Planning over Gaussian Sensor Networks

Emrah Akyol, *Member, IEEE,* and Urbashi Mitra, *Fellow, IEEE*

*Abstract*—Path planning is an important component of autonomous mobile sensing systems. This paper studies upper and lower bounds of communication performance over Gaussian sensor networks, to drive power-distortion metrics for path planning problems. The Gaussian multiple-access channel is employed as a channel model and two source models are considered. In the first setting, the underlying source is estimated with minimum mean squared error, while in the second, reconstruction of a random spatial field is considered. For both problem settings, the upper and the lower bounds of sensor power-distortion curve are derived. For both settings, the upper bounds follow from the amplify-and-forward scheme and the lower bounds admit a unified derivation based on data processing inequality and tensorization property of the maximal correlation measure. Next, closed-form solutions of the optimal power allocation problems are obtained under a weighted sum-power constraint. The gap between the upper and the lower bounds is analyzed for both weighted sum and individual power constrained settings. Finally, these metrics are used to drive a path planning algorithm and the effects of power-distortion metrics, network parameters, and power optimization on the optimized path selection are analyzed.

## I. INTRODUCTION

Sensor networks can provide monitoring and sensing, for surveillance and localization applications as well as scientific studies, over a wide variety of environments such as agricultural fields, desert climes, and underwater systems [2]. In many of these applications, the coverage area of interest is so large that inter-sensor communication is not cost-effective or feasible. For example, in underwater environmental sensing, the ocean is vast and sensors are unlikely to be densely deployed. In these scenarios, it is energy-efficient and cost-effective to employ an autonomous data collecting vehicle (AV) that can travel to all sensors and download the data. In order to communicate with the sensors the AV must be physically close to each sensor, traveling along a path. The *path planning problem* in this context is to find the optimal route along which the AV can collect the data from all sensors at maximum quality with minimal resource use (such as

E. Akyol (akyol@illinois.edu) is with the Coordinated Science Laboratory, University of Illinois at Urbana-Champaign.

U. Mitra (ubli@usc.edu) is with the Electrical Engineering Department, University of Southern California.

This research was funded in part by one or all of these grants: ONR N00014-09-1-0700 AFOSR FA9550-12-1-0215 DOT CA-26-7084-00 NSF CCF-1117896 NSF CNS-1213128 NSF CCF-1410009 NSF CPS-1446901. Part of the material in this paper was presented at the Allerton Conference on Communication, Control and Computing, Urbana, Illinois, Oct. 2014 [1].

overall traveling distance, delay, or total energy spent at the sensors for communication) [3]–[10].

In this paper, we consider lossy reconstruction of the data with minimal distortion and with minimal energy spent at the sensors. More specifically, we derive and analyze the power-distortion characterization of the fundamental limits of communication over sensor networks, with the goal of providing meaningful metrics for robotic path planning. In general, an AV may have two different types of objectives: to estimate the underlying random source, which is typically modeled as a random process in time (referred as source reconstruction, SR, throughout the paper); or to reconstruct the spatial random field generated by the source (field reconstruction, FR).

Consider the motivating example in Fig. 1 which involves one memoryless source and three static sensors. The objective of the AV is to collect source data from the sensors. Each sensor observes the source over a noisy "sensing" channel whose quality depends on the distance between the sensor and the source as follows: if a sensor is closer to the source, it senses the source over a less noisy channel. The sensors transmit their observation to AV over a multiple-access channel. In the first setting of interest, the objective of the AV is to estimate the source signal (SR). It is then intuitively expected that the AV chooses a path toward the sensor(s) which are closest to the source (similar to path (a) in Fig. 1), since they represent the underlying source at maximum fidelity. In the second setting, the objective is to reconstruct the entire spatial random field using the measurements at the sensors. This setting corresponds to a class of environmental monitoring applications where physical quantities, such as the temperature, pressure *etc.* are tracked. In this second case, which we refer as field reconstruction (FR), the objective of the AV simplifies to estimating the sensor measurements, *i.e.,* noisy observations of the source, with minimal weighted distortion, where the weights represent the importance of the associated sensor measurement in reconstructing the spatial field. It is intuitively expected that the optimal AV route would be toward the sensors whose measurements represent the largest field (*i.e.,* the largest weight). In our running example in Fig. 1, the optimal path could look like path (b). In this paper, we formalize these two classes of distortion metrics and analyze the effects of them on the optimized path selection.

We study the power-distortion metrics on a Gaussian sensor network model, see *e.g.,* [11]–[19]. In [12], the performance of a simple amplify-and-forward (AF) scheme is studied, in conjunction with optimal power assignment over the sensors given a sum-power budget. For a particular symmetric setting, Gastpar showed that indeed the AF scheme is optimal over all encoding/decoding methods that allow arbitrarily high delay [20]. However, it was also shown that, in more realistic asymmetric settings, the AF scheme may be suboptimal [21]. In [22], optimal communication strategies were studied for transmitting jointly Gaussian sources over a Gaussian MAC. In general, the optimal communication strategies are *unknown* for both of these settings [23].

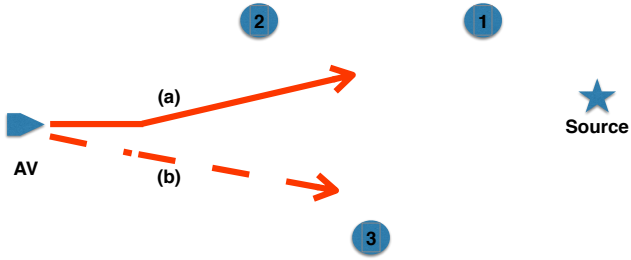

Fig. 1: The underwater data gathering example.

We note that there are other sensor network models beyond the one in this paper, see e.g., [24], and the references therein. Vector settings, associated with the SR metric in our model, were studied in [12] and [19]. Our preliminary results appeared in [1]. The effects of active sensing and adversarial nodes in communications over sensor networks were analyzed in [25] and [26], respectively. Information theoretic analysis of the scaling behavior of such sensor networks was considered in [27].

The contributions of this paper are the following:

- Building on [22] and [20], we derive the lower and the upper bounds of the power-distortion functions for the two problem settings. The upper bounds are obtained through the AF scheme, while the lower bounds follow from the data processing inequality used in conjunction with the tensorization property of the maximal correlation, also known as the Witsenhausen's lemma [28].
- For each of these metrics, we derive, in closed-form, strategies for optimal the power allocation over sensors subject to a weighted-sum power constraint. Perhaps surprisingly, in all cases, the original problem is non-convex in the individual power allocation variables, however it can be translated into a convex optimization problem that admits a closed-form solution.
- We provide numerical analysis of the different metrics and uncover cases where the gap between the upper and the lower bounds is small (or large), implying the near optimality of the AF scheme (or the need for more sophisticated coding schemes). Our results associated with the SR metric imply that when the sensing and the communication channels are matched- *i.e*, the sensor with better sensing channel has the better communication channel- the performance loss due to using the AF scheme is low. On the other hand, this loss increases when the sensing and the communication channels are inversely matched, *i.e.,* when the sensor with the better sensing channel has the worse communication channel. However, for the FR metric, the AF scheme performs significantly worse than the lower bound. We also observe that in general, the lower bounds are more sensitive to the choice of parameters than the upper bounds.

- Based on the proposed metrics, we implement a simple path planning algorithm. We analyze, via numerical simulations, the impact of power optimization and metric selection on the robustness of the path planning to the sensing/communication channel parameters and the network topology.

This paper is organized as follows: we present the communication model along with the specific metrics in Section II. We present our results regarding lower and upper bounds of communication performance with and without power optimization in Section III. We numerically analyze these metrics and their use in path planning in Section IV. We present conclusions and discussion in Section V.

## II. COMMUNICATION MODEL

### A. Notation

Let $\mathbb{E}(\cdot)$ and $|| \cdot ||_2$ denote the expectation and $l_2$ norm operators, and $\mathbb{R}$ and $\mathbb{R}^+$ denote the set of real and positive real numbers, respectively. In general, lowercase letters (*e.g.*, $x$) denote scalars, boldface lowercase (*e.g.*, $\boldsymbol{x}$) vectors, uppercase (*e.g.*, $U, X$) matrices and random variables, and boldface uppercase (*e.g.*, $\boldsymbol{X}$) random vectors. Unless otherwise specified, vectors and random vectors have length $m$, and matrices have size $m \times m$. The $k^{th}$ element of vector $\boldsymbol{x}$ is denoted by $[\boldsymbol{x}]_k$ and the $(i,j)^{th}$ element and the $k^{th}$ column of the matrix $A$ are denoted by $[A]_{ij}$ and $[A]_k$ respectively. Let $A^T$ denote the transpose of matrix $A$. A diagonal matrix with diagonal elements $\boldsymbol{a}$ is denoted by $\text{diag}(\boldsymbol{a})$. $R_X$ and $R_{XZ}$ denote the auto-covariance of $\boldsymbol{X}$ and cross covariance of $\boldsymbol{X}$ and $\boldsymbol{Z}$ respectively. Gaussian distribution with mean $\boldsymbol{\mu}$ and covariance matrix $R$ is denoted as $\mathcal{N}(\boldsymbol{\mu}, R)$. The mutual information between random variables $X$ and $Y$ is denoted as $I(X;Y)$.

### B. Problem Definition

The problem setting is depicted in Fig. 2 where the underlying source $\{S(i)\}$ is a sequence of independently and identically distributed (i.i.d.) real valued Gaussian random variables with zero mean and unit variance, without loss of generality. We consider a pre-deployed network of $M$ sensors. Sensor $m$ observes a sequence $\{U_m(i)\}$ defined as

$$U_m(i) = \beta_m S(i) + W_m(i), \tag{1}$$

where $\{W_m(i)\}$ is a sequence of i.i.d. Gaussian random variables with zero mean and unit variance, independent of $\{S(i)\}$. Sensor $m$ applies the encoding function $g_m^N : \mathbb{R}^N \to \mathbb{R}^N$ to the observation sequence of length $N$, $\boldsymbol{U}_m$ to generate a sequence of channel inputs $X_m = g_m^N(\boldsymbol{U}_m)$ that satisfies

$$\frac{1}{N} \sum_{i=1}^N \mathbb{E}\{X_m^2(i)\} \leq P_m \tag{2}$$

where $P_m$ is the individual power constraint on sensor $m$. This problem formulation presumes *fixed* power budget, $P_m$,

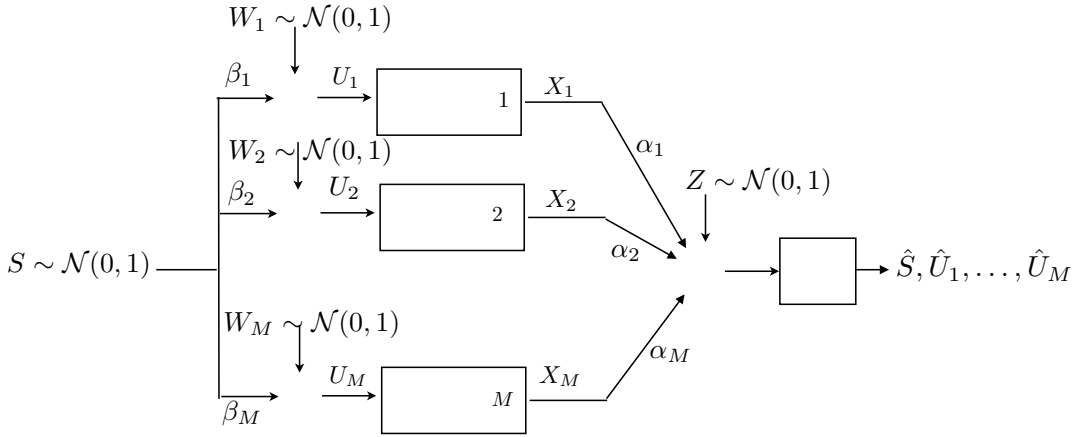

Fig. 2: The sensor network model

for each sensor. Another problem setting we consider involves a weighted sum-power constraint in the form of

$$\frac{1}{N} \sum_{m=1}^{M} r_m \sum_{i=1}^{N} \mathbb{E}\{X_m^2(i)\} \leq P_T, \tag{3}$$

where the weight vector $\boldsymbol{r} = [r_1, r_2 \ldots r_M]$ is known. The channel output is

$$Y(i) = Z(i) + \sum_{m=1}^{M} \alpha_m X_m(i), \tag{4}$$

where $\{Z(i)\}$ is a sequence of i.i.d. Gaussian random variables of zero mean and unit variance, independent of $\{S(i)\}$ and $\{W_m(i)\}$. The receiver applies a function $h^N : \mathbb{R}^N \to \mathbb{R}^N$ to the received sequence $\{Y\}$ to minimize the cost-which is defined explicitly in the next section for each scenario of interest.

With a slight abuse of notation, we let $J(\boldsymbol{P})$ denote the distortion metric with the power allocation vector $\boldsymbol{P} = [P_1, P_2, \ldots, P_M]$ (assigned power to the sensor $m$ is denoted by $P_m$), and $J(P_T)$ denote the metric with total power $P_T$ with optimized power allocation. The sensor network parameters, $\boldsymbol{\beta}$, and the communication channel parameters, $\boldsymbol{\alpha}$, are fixed and known to the sensors and the receiver; and the block-length $N$ is asymptotically large, *i.e.*, $N \to \infty$.

## III. DISTORTION METRICS

### A. Source reconstruction

The source reconstruction (SR) metric, denoted as $J_S$, involves estimating the underlying source $S$ with minimum mean squared error (MSE):

$$J_S(\boldsymbol{P}) \triangleq \lim_{N \to \infty} \frac{1}{N} \sum_{i=1}^{N} \mathbb{E}\{(S(i) - \hat{S}(i))^2\}, \tag{5}$$

where $\hat{S}(i)$ is the estimate of $S(i)$ at the receiver. While the exact characterization of $J_S(\boldsymbol{P})$ is in general difficult [23], we state upper and lower bounds of $J_S(\boldsymbol{P})$ in the following theorem.

**Theorem 1.** *For any given* $\boldsymbol{P}$, $J_S^L(\boldsymbol{P}) \leq J_S(\boldsymbol{P}) \leq J_S^U(\boldsymbol{P})$ *holds where* $J_S^U(\boldsymbol{P})$ *and* $J_S^L(\boldsymbol{P})$ *are given in (6) and (7) respectively.*

*Proof:* The derivation of $J_S^U(\boldsymbol{P})$ follows directly from the AF communication scheme, where each sensor scales its input $U_m(i)$, symbol-by-symbol, to match $\mathbb{E}\{X_m^2(i)\}$ to the allowed power $P_m$ for each time instant $i$, *i.e.*, $X_m(i) = \sqrt{\frac{P_m}{1+\beta_m^2}} U_m(i)$. We have

$$J_S^U(\boldsymbol{P}) = \mathbb{E}\{S^2\} - \mathbb{E}\{SY\}(\mathbb{E}\{Y^2\})^{-1}\mathbb{E}\{YS\} \tag{8}$$

where

$$\mathbb{E}\{SY\} = \sum_{m=1}^{M} \beta_m \alpha_m \sqrt{\frac{P_m}{1+\beta_m^2}}, \tag{9}$$

and

$$\mathbb{E}\{Y^2\} = 1 + \left( \sum_{m=1}^{M} \beta_m \alpha_m \sqrt{\frac{P_m}{1+\beta_m^2}} \right)^2 + \sum_{m=1}^{M} \alpha_m^2 \frac{P_m}{1+\beta_m^2}. \tag{10}$$

Plugging (9) and (10) into (8), we obtain (6).

For $J_S^L(\boldsymbol{P})$, we follow the steps in [20] to generalize its main result for the symmetric setting to the asymmetric setting considered here. First, we note that from the data processing theorem, we must have

$$I(\boldsymbol{U}_1, \boldsymbol{U}_2, \ldots, \boldsymbol{U}_M; \hat{\boldsymbol{S}}) \leq I(\boldsymbol{X}_1, \boldsymbol{X}_2, \ldots, \boldsymbol{X}_M; \boldsymbol{Y}) \tag{11}$$

The left hand side can be lower bounded as:

$$I(\boldsymbol{U}_1, \boldsymbol{U}_2, \ldots, \boldsymbol{U}_M; \hat{\boldsymbol{S}}) \geq NR(D) \tag{12}$$

where $R(D)$ is derived in the Appendix A. The right hand side can be upper bounded by

$$I(\boldsymbol{X}_1, \ldots, \boldsymbol{X}_M; \boldsymbol{Y}) \leq \sum_{i=1}^{N} I(X_1(i), \ldots, X_M(i); Y(i)) \tag{13}$$

$$\leq \max \sum_{i=1}^{N} I(X_1(i), \ldots, X_M(i); Y(i)) \tag{14}$$

$$= \frac{1}{2} \sum_{i=1}^{N} \log(1 + \boldsymbol{\alpha}^T R_X(i)\boldsymbol{\alpha}) \tag{15}$$

$$J_S^U(\boldsymbol{P}) = \frac{1 + \sum\limits_{m=1}^{M} \alpha_m^2 \frac{P_m}{1+\beta_m^2}}{1 + \left(\sum\limits_{m=1}^{M} \beta_m \alpha_m \sqrt{\frac{P_m}{1+\beta_m^2}}\right)^2 + \sum\limits_{m=1}^{M} \alpha_m^2 \frac{P_m}{1+\beta_m^2}}, \tag{6}$$

$$J_S^L(\boldsymbol{P}) = \frac{1}{1 + \sum\limits_{m=1}^{M} \beta_m^2} \left(1 + \frac{\sum\limits_{m=1}^{M} \beta_m^2}{1 + \sum\limits_{m=1}^{M} \alpha_m^2 \frac{P_m}{1+\beta_m^2} + \left(\sum\limits_{m=1}^{M} \beta_m \alpha_m \sqrt{\frac{P_m}{1+\beta_m^2}}\right)^2}\right). \tag{7}$$

where $R_X(i)$ is defined as

$$\{R_X(i)\}_{p,r} \triangleq \mathbb{E}\{X_p(i)X_r(i)\} \quad \forall p, r \in [1:M]. \tag{16}$$

Note that (13) follows from the memoryless property of the channel and the maximum in (14) is over the joint density of $X_1(i), \ldots, X_{M+K}(i)$, given the structural constraints on $R_X(i)$ due to the power constraints. It is well known that the maximum is achieved by the jointly Gaussian density for a given covariance, yielding (15). Since the logarithm is a monotonically increasing function, the optimal encoding functions $g_m^N(\cdot), m \in [1:M]$ equivalently maximize $\sum\limits_{p,r} \mathbb{E}\{X_p(i)X_r(i)\}$ for all $i$. Note that

$$X_m = g_m^N(\boldsymbol{U}_m) \tag{17}$$

and hence the $g_m^N(\cdot), m \in [1:M]$ that maximize $\sum\limits_{p,r} \mathbb{E}\{X_p(i)X_r(i)\}$ can be found by invoking Witsenhausen's lemma (given in Appendix C) as $X_m(i) = \boldsymbol{\gamma}_N \boldsymbol{U}_m$ where

$$\boldsymbol{\gamma}_N = \left(\sqrt{\frac{P_m}{1+\beta_m^2}}, \ldots, \sqrt{\frac{P_m}{1+\beta_m^2}}\right). \tag{18}$$

Plugging (18) in (12), we have

$$R = \frac{1}{2} \log \left(1 + \sum\limits_{m=1}^{M} \alpha_m^2 \frac{P_m}{1+\beta_m^2} + \left(\sum\limits_{m=1}^{M} \beta_m \alpha_m \sqrt{\frac{P_m}{1+\beta_m^2}}\right)^2\right) \tag{19}$$

Plugging the expression for $R(D)$ (derived in the Appendix A) in (75), we obtain (7). ∎

**Remark 1.** *It is of interest to see whether the mutual information or source SNR based metrics (see e.g., [29], [30]) are of use here. Noting that $S$ and $\hat{S}$ used in the derivation of both lower and upper bounds are jointly Gaussian, it is straightforward to show that $I(S; \hat{S}) = -\frac{1}{2} \log J_S(\boldsymbol{P})$. Hence, minimizing $J_S(\boldsymbol{P})$ and maximizing $I(S; \hat{S})$ are effectively identical for path selection purposes. The same conclusion also holds for the field reconstruction metric (defined in the next section) by similar arguments. Also note that the source SNR is exactly $1/J_S(P)$ and hence, maximizing source SNR is equivalent to minimizing $J_S(P)$.*

Next, we discuss the optimal power allocation among sensors and derive the optimal trade-off between distortion metrics and the weighted sum of transmit power. Particularly,

we study the following optimization problem:

$$\underset{P_1, P_2, \ldots, P_M}{\text{minimize}} \quad J_S(P_1, P_2, \ldots, P_M)$$

$$\text{subject to} \quad \sum_{m=1}^{M} r_m P_m \leq P_T$$

where $\boldsymbol{r} = [r_1, r_2, \ldots, r_M]$ and $P_T$ are given optimization parameters. The following theorem states the $P_T$ versus $J_S$ relationship when the power allocation is optimized.

**Theorem 2.** *For any given $P_T$ and a weight vector $\boldsymbol{r} = [r_1, r_2 \ldots r_M]$, $J_S^L(P_T) \leq J_S(P_T) \leq J_S^U(P_T)$ holds where*

$$J_S^U(P_T) = \left(1 + P_T \sum_{m=1}^{M} \frac{\alpha_m^2 \beta_m^2}{r_m + r_m \beta_m^2 + P_T \alpha_m^2}\right)^{-1} \tag{20}$$

$$J_S^L(P_T) = \frac{1}{1 + \sum\limits_{m=1}^{M} \beta_m^2} \left(1 + \frac{\sum\limits_{m=1}^{M} \beta_m^2}{1 + \frac{P_T}{\lambda}}\right) \tag{21}$$

*and $\lambda$ satisfies*

$$\sum_{m=1}^{M} \frac{\alpha_m^2 \beta_m^2}{r_m + r_m \beta_m^2 - \lambda \alpha_m^2} = \frac{1}{\lambda}. \tag{22}$$

*Proof:* $J_S^U(P_T)$: The proof follows from similar steps of the proof of Theorem 4 of [12] with appropriate changes due to the weight vector $\boldsymbol{r}$, and is omitted.

$J_S^L(P_T)$: Minimization of $J_S^L(\boldsymbol{P})$ in $\boldsymbol{P}$ is equivalent to minimizing

$$D = -\left(\sum_{m=1}^{M} \beta_m \alpha_m \sqrt{\frac{P_m}{1+\beta_m^2}}\right)^2 - \sum_{m=1}^{M} \alpha_m^2 \frac{P_m}{1+\beta_m^2} \tag{23}$$

subject to $\sum\limits_{m=1}^{M} r_m P_m \leq P_T$ over $P_m \geq 0$ for all $m$. This objective function is not convex[1] in the variables $P_m$. We first impose a slackness variable

$$t = \sum_{m=1}^{M} \alpha_m \beta_m \sqrt{\frac{P_m}{1+\beta_m^2}}. \tag{24}$$

and analyze the dual problem: minimize

$$\sum_{m=1}^{M} r_m P_m, \tag{25}$$

---

[1]This can easily be shown by checking the positive definiteness Hessian of the objective function.

subject to

$$-D - \sum_{m=1}^{M} \alpha_m^2 \frac{P_m}{1+\beta_m^2} \le t^2, \qquad (26)$$

and (24). This problem *is* convex in the variables $P_m$ and $t$, and can be converted into unconstrained optimization problem: minimize

$$J = \sum_{m=1}^{M} r_m P_m + \lambda_1 \left( -D - \sum_{m=1}^{M} \alpha_m^2 \frac{P_m}{1+\beta_m^2} - t^2 \right)$$
$$+ \lambda_2 \left( t - \sum_{m=1}^{M} \alpha_m \beta_m \sqrt{\frac{P_m}{1+\beta_m^2}} \right), \quad (27)$$

where $\lambda_1 \in \mathbb{R}^+$ and $\lambda_2 \in \mathbb{R}$. Next, we note that Karush-Kuhn-Tucker (KKT) optimality conditions are sufficient for this problem (no duality gap) since the objective function is convex in the variables $P_m$ and $t$ [31]. We determine the KKT conditions:

$$\frac{\partial J}{\partial P_m} = r_m - \lambda_1 \frac{\alpha_m^2}{1+\beta_m^2} - \lambda_2 \frac{\alpha_m \beta_m}{2\sqrt{P_m(1+\beta_m^2)}} = 0, \quad (28)$$

$$\frac{\partial J}{\partial t} = -2\lambda_1 t + \lambda_2 = 0, \qquad (29)$$

$$-D - \sum_{m=1}^{M} \alpha_m^2 \frac{P_m}{1+\beta_m^2} = t^2, \qquad (30)$$

and we have (24). From (28), we obtain $P_m$ in terms of $\lambda_1$ and $\lambda_2$ as

$$P_m = \frac{\lambda_2^2}{4} \frac{\alpha_m^2 \beta_m^2 (1+\beta_m^2)}{(r_m + r_m \beta_m^2 - \lambda_1 \alpha_m^2)^2}. \qquad (31)$$

Plugging (31) into (24), we have

$$t = \frac{\lambda_2}{2} \sum_{m=1}^{M} \frac{\alpha_m^2 \beta_m^2}{r_m + r_m \beta_m^2 - \lambda_1 \alpha_m^2}. \qquad (32)$$

Re-writing (30) using (29) and (32), we have

$$\frac{\lambda_2^2}{4\lambda_1} \sum_{m=1}^{M} \frac{\alpha_m^2 \beta_m^2}{r_m + r_m \beta_m^2 - \lambda_1 \alpha_m^2} = -D$$
$$- \frac{\lambda_2^2}{4} \sum_{m=1}^{M} \frac{\alpha_m^4 \beta_m^2}{(r_m + r_m \beta_m^2 - \lambda_1 \alpha_m^2)^2} \qquad (33)$$

which simplifies to

$$-D = \frac{\lambda_2^2}{4\lambda_1} \sum_{m=1}^{M} \frac{r_m(1+\beta_m^2)\alpha_m^2 \beta_m^2}{(r_m + r_m \beta_m^2 - \lambda_1 \alpha_m^2)^2} \qquad (34)$$

$$= \frac{1}{\lambda_1} \sum_{m=1}^{M} r_m P_m = \frac{P_T}{\lambda_1}. \qquad (35)$$

Expressing (32) using (29), we obtain (22). Using (35), we obtain (21). ∎

**Remark 2.** *The coefficient $\lambda$ is a Lagrange parameter in a convex optimization problem, as demonstrated in the proof above; hence, the solution of (22) exists and it is unique [31]. It can be found numerically by a bisection search. In practice, the computational complexity of determining $\lambda$ is relatively low, since it is computed only once for each network setting, i.e., it does not depend on $P_T$.*

### B. Field Reconstruction

In the field reconstruction (FR) setting, the objective of the receiver is to estimate the entire random field which is covered by the sensors. We assume that at any point is represented by the closest sensor, or alternatively a linear interpolation of the closest $k$ sensors readings. Hence, we define the FR metric as

$$J_F(\boldsymbol{P}) \triangleq \lim_{N \to \infty} \frac{1}{N} \sum_{i=1}^{N} \sum_{m=1}^{M} \gamma_m \mathbb{E}\{(U_m(i) - \hat{U}_m(i))^2\} \quad (36)$$

where the $\gamma_m$ are determined by $k$ and network parameters (*i.e.*, sensor locations). Before stating the results, we define the covariance matrix of sensor inputs $U$, i.e., $R_U \triangleq \mathbb{E}\{\boldsymbol{U}\boldsymbol{U}^T\}$ which can explicitly be expressed as a function of $\boldsymbol{\beta}$

$$R_U = \begin{pmatrix} 1+\beta_1^2 & \beta_1\beta_2 & \dots & \beta_1\beta_M \\ \beta_1\beta_2 & 1+\beta_2^2 & \dots & \beta_2\beta_M \\ \vdots & & \ddots & \vdots \\ \beta_1\beta_M & \dots & & 1+\beta_M^2 \end{pmatrix}. \qquad (37)$$

$R_U$ admits an eigen-decomposition $R_U = Q_U^T \Lambda Q_U$ where $Q_U$ is unitary and $\Lambda$ is a diagonal matrix with elements $1, \dots, 1, 1+\sum_m \beta_m^2$. The following *transformed* weight vector is used in the subsequent results:

$$\boldsymbol{\gamma}_k' \triangleq [Q_U^T \operatorname{diag}(\boldsymbol{\gamma})Q_U]_{kk}. \qquad (38)$$

Again, the complete characterization of $J_F(\boldsymbol{P})$ is difficult in general [23], and similar to the SR case, we state upper and lower bounds in the following theorem. In the derivation of $J_F^U(\boldsymbol{P})$, we assume a high power (low distortion) regime, in order to simplify results.

**Theorem 3.** *For a given $\boldsymbol{P}$, $J_F^L(\boldsymbol{P}) \le J_F(\boldsymbol{P}) \le J_F^U(\boldsymbol{P})$ holds, where $J_F^U(\boldsymbol{P})$ and $J_F^L(\boldsymbol{P})$ are given in (39) and (40) respectively. where $A = \sum_{m=1}^{M} \beta_m \alpha_m \sqrt{\frac{P_m}{1+\beta_m^2}}$.*

*Proof:* The derivation of $J_F^U(\boldsymbol{P})$ follows from the AF scheme, where for each $m$, we have

$$J_m = \mathbb{E}\{U_m^2\} - \mathbb{E}\{U_m Y\}(\mathbb{E}\{Y^2\})^{-1}\mathbb{E}\{YU_m\} \qquad (41)$$

Noting that

$$\mathbb{E}\{U_m Y\} = \alpha_m \sqrt{\frac{P_m}{1+\beta_m^2}} + \beta_m \sum_{m=1}^{M} \beta_m \alpha_m \sqrt{\frac{P_m}{1+\beta_m^2}}, \quad (42)$$

and using the expression for $(\mathbb{E}\{Y^2\})^{-1}$ given in (10), we have:

$$J_m = 1+\beta_m^2 - \frac{\left(\alpha_m \sqrt{\frac{P_m}{1+\beta_m^2}} + \beta_m \sum_{m=1}^{M} \beta_m \alpha_m \sqrt{\frac{P_m}{1+\beta_m^2}}\right)^2}{1 + \left(\sum_{m=1}^{M} \beta_m \alpha_m \sqrt{\frac{P_m}{1+\beta_m^2}}\right)^2 + \sum_{m=1}^{M} \alpha_m^2 \frac{P_m}{1+\beta_m^2}}. \quad (43)$$

Noting that $J_F^U(\boldsymbol{P}) = \sum_{m=1}^{M} \gamma_m J_m$, we obtain (39). To derive $J_F^L(\boldsymbol{P})$, we follow the steps (11)-(19) *verbatim*, with the difference that we use vector $R(D)$ (derived in Appendix B) instead of $R(D)$ associated with remote compression. Combining (19) with (84), we obtain (40). ∎

$$J_F^U(\boldsymbol{P}) = -\frac{\sum\limits_{m=1}^{M} \gamma_m \alpha_m^2 \frac{P_m}{1+\beta_m^2} + A^2 \left(\sum\limits_{n=1}^{M} \gamma_m \beta_m^2\right) + 2A \left(\sum\limits_{m=1}^{M} \gamma_m \beta_m \alpha_m \sqrt{\frac{P_m}{1+\beta_m^2}}\right)}{1 + A^2 + \sum\limits_{m=1}^{M} \alpha_m^2 \frac{P_m}{1+\beta_m^2}} + \sum\limits_{m=1}^{M} \gamma_m(1+\beta_m^2) \tag{39}$$

$$J_F^L(\boldsymbol{P}) = M \left(\left(1+\sum\limits_{m=1}^{M}\beta_m^2\right)\prod\limits_{m=1}^{M}\gamma_m'\right)^{\frac{1}{M}} \left(1+\sum\limits_{m=1}^{M}\alpha_m^2 \frac{P_m}{1+\beta_m^2} + A^2\right)^{-\frac{1}{M}} \tag{40}$$

**Remark 3.** *As a side note, by (76), we note that $\boldsymbol{\gamma}$ majorizes $\boldsymbol{\gamma}'$ (see [32, 2.B.3]) which implies that $\prod\limits_{m=1}^{M}\gamma_m' \geq \prod\limits_{m=1}^{M}\gamma_m$. A looser bound (independent of $\boldsymbol{\gamma}'$) can be obtained by replacing $\boldsymbol{\gamma}'$ with $\boldsymbol{\gamma}$.*

Next, we optimize power allocation over the sensors. For notational convenience, here we assume[2] $\gamma_m = 1$ for all $m$. The following theorem states the upper and lower bounds of the FR metric with power optimization.

**Theorem 4.** *For any given $P_T$ and $\boldsymbol{r} = [r_1, r_2 \ldots r_M]$, $J_F^L(P_T) \leq J_F(P_T) \leq J_F^U(P_T)$ holds where*

$$J_F^U(P_T) = M + \sum\limits_{m=1}^{M}\beta_m^2 + \frac{P_T}{\lambda_1 - P_T}, \tag{44}$$

$$J_F^L(P_T) = M \left(\left(1+\sum\limits_{m=1}^{M}\beta_m^2\right)\frac{\lambda_2}{P_T}\right)^{\frac{1}{M}}, \tag{45}$$

*and $\lambda_1$ and $\lambda_2$ satisfy*

$$\left(-1-\left(1-\frac{P_T}{\lambda_1}\right)\left(1+\sum\limits_{m=1}^{M}\beta_m^2\right)\right. \\ \left.\times \sum\limits_{m=1}^{M}\frac{\alpha_m^2\beta_m^2}{r_m+r_m\beta_m^2+\lambda_1\alpha_m^2}\right) = \frac{1}{\lambda_1}, \tag{46}$$

$$\sum\limits_{m=1}^{M}\frac{\alpha_m^2\beta_m^2}{r_m+r_m\beta_m^2-\lambda_2\alpha_m^2} = \frac{1}{\lambda_2}. \tag{47}$$

*Proof:* Plugging $\gamma_m = 1$ for all $m$ in (39), we have (48). Minimization of $J_F^U(\boldsymbol{P})$ in $\boldsymbol{P}$ is equivalent to minimizing

$$D = -\frac{\sum\limits_{m=1}^{M}\alpha_m^2\frac{P_m}{1+\beta_m^2} + \left(2+\sum\limits_{m=1}^{M}\beta_m^2\right)\left(\sum\limits_{m=1}^{M}\beta_m\alpha_m\sqrt{\frac{P_m}{1+\beta_m^2}}\right)^2}{1+\left(\sum\limits_{m=1}^{M}\beta_m\alpha_m\sqrt{\frac{P_m}{1+\beta_m^2}}\right)^2 + \sum\limits_{m=1}^{M}\alpha_m^2\frac{P_m}{1+\beta_m^2}} \tag{49}$$

subject to

$$\sum\limits_{m=1}^{M} r_m P_m \leq P_T \tag{50}$$

for $P_m \geq 0$ for all $m$. This objective function is not convex in $P_m$. Following similar steps to those in the proof of Theorem

2, we first convert the problem into a convex form introducing a slack variable

$$t = \sum\limits_{m=1}^{M}\alpha_m\beta_m\sqrt{\frac{P_m}{1+\beta_m^2}}, \tag{51}$$

and express the optimizing problem as to minimize $\sum\limits_{m=1}^{M} r_m P_m$ subject to

$$\frac{D}{1+D} + \sum\limits_{m=1}^{M}\alpha_m^2\frac{P_m}{1+\beta_m^2} \leq \delta t^2, \tag{52}$$

and (51) where

$$\delta \triangleq -1 - \frac{1+\sum\limits_{m=1}^{M}\beta_m^2}{1+D}. \tag{53}$$

This problem *is* convex in the variables $P_m$ and $t$, hence can be converted into an unconstrained optimization problem where we minimize

$$J = \sum\limits_{m=1}^{M} r_m P_m + \lambda_1 \left(\frac{D}{1+D} + \sum\limits_{m=1}^{M}\alpha_m^2\frac{P_m}{1+\beta_m^2} - \delta t^2\right) \tag{54}$$

$$+ \lambda_2 \left(t - \sum\limits_{m=1}^{M}\alpha_m\beta_m\sqrt{\frac{P_m}{1+\beta_m^2}}\right), \tag{55}$$

where $\lambda_1 \in \mathbb{R}^+$ and $\lambda_2 \in \mathbb{R}$. Applying the KKT conditions, we have

$$\frac{\partial J}{\partial P_m} = r_m + \lambda_1 \frac{\alpha_m^2}{1+\beta_m^2} - \lambda_2 \frac{\alpha_m\beta_m}{2\sqrt{P_m(1+\beta_m^2)}} = 0, \tag{56}$$

$$\frac{\partial J}{\partial t} = -2\lambda_1\delta t + \lambda_2 = 0, \tag{57}$$

$$\frac{\partial J}{\partial \lambda_1} = \frac{D}{1+D} + \sum\limits_{m=1}^{M}\alpha_m^2\frac{P_m}{1+\beta_m^2} - \delta t^2 = 0. \tag{58}$$

From (56), we obtain $P_m$ in terms of $\lambda_1$ and $\lambda_2$ as

$$P_m = \frac{\lambda_2^2}{4}\frac{\alpha_m^2\beta_m^2(1+\beta_m^2)}{(r_m+r_m\beta_m^2+\lambda_1\alpha_m^2)^2}. \tag{59}$$

Plugging (59) in (51), we have

$$t = \frac{\lambda_2}{2}\sum\limits_{m=1}^{M}\frac{\alpha_m^2\beta_m^2}{r_m+r_m\beta_m^2+\lambda_1\alpha_m^2}. \tag{60}$$

Re-writing (58) using (57) and (60), we have (61) which

---

[2]Note that this assumption does not introduce any loss of generality, since $\boldsymbol{\gamma}$ can be incorporated into the power scaling coefficients $\boldsymbol{\beta}$ in the problem definition.

$$J_F^U(\boldsymbol{P}) = M + \sum_{m=1}^{M} \beta_m^2 - \frac{\sum_{m=1}^{M} \alpha_m^2 \frac{P_m}{1+\beta_m^2} + \left(2 + \sum_{m=1}^{M} \beta_m^2\right)\left(\sum_{m=1}^{M} \beta_m \alpha_m \sqrt{\frac{P_m}{1+\beta_m^2}}\right)^2}{1 + \left(\sum_{m=1}^{M} \beta_m \alpha_m \sqrt{\frac{P_m}{1+\beta_m^2}}\right)^2 + \sum_{m=1}^{M} \alpha_m^2 \frac{P_m}{1+\beta_m^2}} \tag{48}$$

$$\frac{\lambda_2^2}{4\lambda_1} \sum_{m=1}^{M} \frac{\alpha_m^2 \beta_m^2}{r_m + r_m \beta_m^2 + \lambda_1 \alpha_m^2} = \frac{D}{1+D} + \frac{\lambda_2^2}{4} \sum_{m=1}^{M} \frac{\alpha_m^4 \beta_m^2}{\left(r_m + r_m \beta_m^2 + \lambda_1 \alpha_m^2\right)^2} \tag{61}$$

simplifies to

$$\lambda_1 \frac{D}{1+D} = \frac{\lambda_2^2}{4} \sum_{m=1}^{M} \frac{r_m(1+\beta_m^2)\alpha_m^2 \beta_m^2}{\left(r_m + r_m \beta_m^2 + \lambda_1 \alpha_m^2\right)^2} \tag{62}$$

$$= \sum_{m=1}^{M} r_m P_m = P_T. \tag{63}$$

Plugging (63) in (48), we obtain $J_F^U(P_T)$. Plugging (53), (57), and (63) in (60), we obtain (46).

$J_F^L(P_T)$: Minimization of $J_F^L(\boldsymbol{P})$ in $\boldsymbol{P}$ is equivalent to minimizing

$$D = -\left(\sum_{m=1}^{M} \beta_m \alpha_m \sqrt{\frac{P_m}{1+\beta_m^2}}\right)^2 - \sum_{m=1}^{M} \alpha_m^2 \frac{P_m}{1+\beta_m^2}, \tag{64}$$

subject to $\sum_{m=1}^{M} r_m P_m \leq P_T$ and $P_m \geq 0$ for all $m$. This problem is solved in the proof of Theorem 2, hence we follow the same steps as in (24)-(35) and obtain $J_F^L(P_T)$. ∎

**Remark 4.** *Similar to the SR setting (see Remark 2), $\lambda_1$ and $\lambda_2$ in (47) and (46) are in fact Lagrange parameters in a convex optimization problem as shown in the proof above, hence they exist and they are unique [31]. Unlike (22), the computation of $\lambda_1$ in (46) depends on $P_T$ in addition to $\boldsymbol{\alpha}$ and $\boldsymbol{\beta}$. However, for a given $P_T$, the computation employs similar steps.*

**Remark 5.** *The optimal power allocation strategies in both SR and FR settings can be implemented by each sensor in a distributed manner: the central agent (e.g., the AV) computes the optimal values of $\lambda_1$ and $\lambda_2$ (or $\lambda$ in SR setting), and broadcasts this information to all sensors. Each sensor then computes its own power allocation based on local parameters $\alpha_m$ and $\beta_m$ and the broadcasted global parameters $\lambda_1$ and $\lambda_2$ (or $\lambda$).*

## IV. NUMERICAL RESULTS

We first analyze different metrics, particularly the gap between upper and lower bounds and the impact of power optimization. Next, we focus on the problem of path planning in conjunction with these metrics.

### A. Metrics

In our experiments, we select $\boldsymbol{\alpha}$ and $\boldsymbol{\beta}$ randomly, uniformly from the interval $[0, 1]$. To analyze the impact of sensing and communication channel ordering on the metrics, we look at

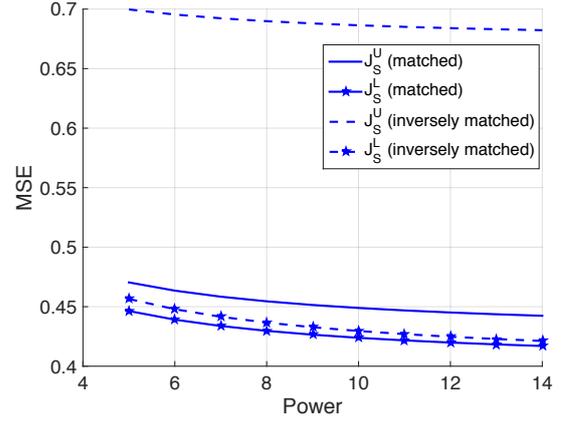

(a) Comparison of MSE bounds for matched and mismatched channels for source reconstruction.

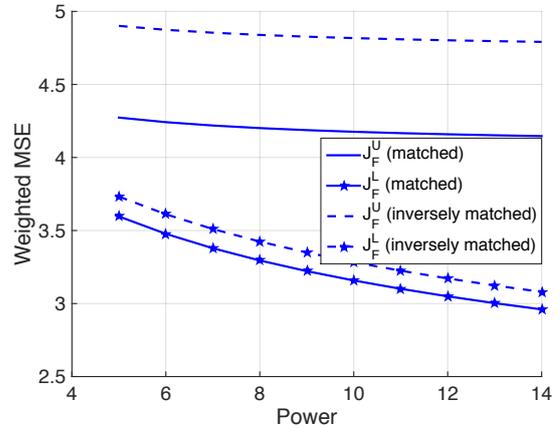

(b) Comparison of MSE bounds for matched and mismatched channels for field reconstruction.

Fig. 3: MSE bounds for uniform power allocation.

two extreme cases: i) ordered channels, *i.e.*, the better sensing channel (larger $\beta$) is matched to better communication channel (larger $\alpha$), ii) reverse ordered, *i.e.*, larger $\beta$ is matched to smaller $\alpha$. For the FR metric, we set $\gamma_m = 1$ for all $m$. To obtain statistically meaningful results, we average the results over 10000 runs of this experiment. We set the number of sensors 5, i.e., $M = 5$.

In Fig. 3, we plot the comparative results with individual power constraints. All sensors are assumed to have identical power, $P_m = P$ for all $m$. As can be seen numerically, when the sensing and the communication channels are matched, *i.e.*, the sensor with the better sensing channel sees a better

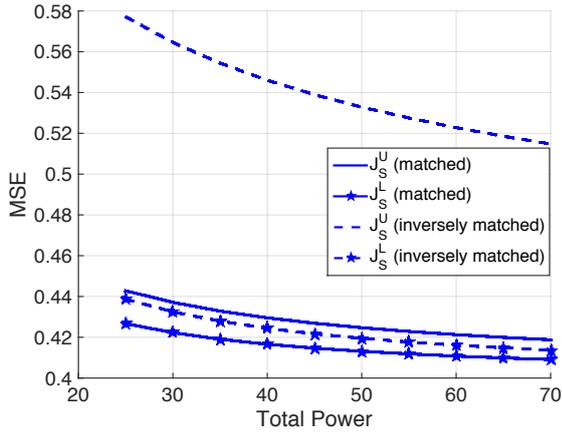

(a) Comparison of MSE bounds for matched and mismatched channels for source reconstruction.

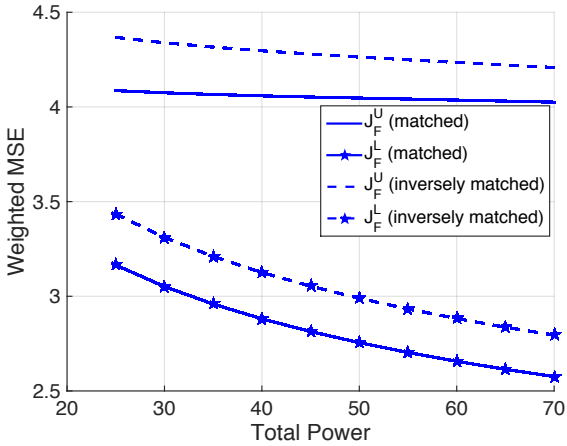

(b) Comparison of MSE bounds for matched and mismatched channels for field reconstruction.

Fig. 4: MSE bounds for optimized power allocation.

communication channel, the gap between upper and lower bounds is small, and as they get mismatched, this gap widens. An important observation is that in the matched order case, upper and lower bounds perform very close for both settings, particularly for the SR setting.

In Fig. 4, we plot the comparative results with a total power constraint, with $r_m = 1$ for all $m$.

### B. Path Planning Results

To obtain the optimal paths given these metrics, we use a simple search algorithm based on determining the step (in four directions) at each point in a greedy manner, i.e., the AV at location $(i, j)$ moves to $(i \pm 1, j)$ or $(i, j \pm 1)$ or stays at $(i, j)$ depending on the cost at these locations. More sophisticated search algorithms can be found in the robotics literature (see *e.g.,* [7] and the references therein). We note that the optimal path will depend on the specific search algorithm used. Our objective here is to demonstrate the use of the proposed metrics in path planning, and their implications on the chosen path, i.e., the type of distortion metric chosen and network parameters affect the optimal path significantly. Our numerical examples demonstrate this conclusion via numerical examples generated using this simple search algorithm.

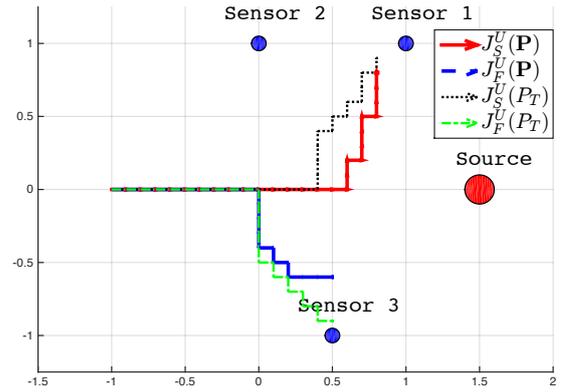

(a) Paths driven by upper bounds

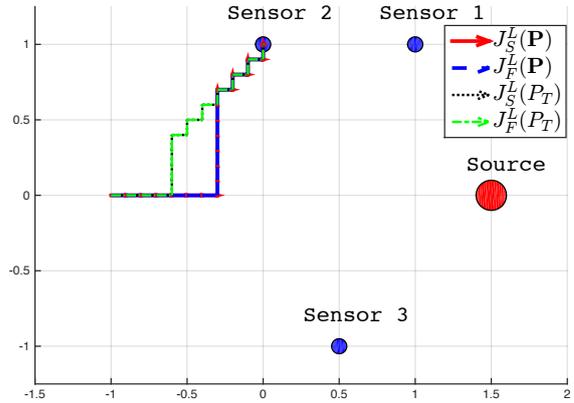

(b) Paths driven by lower bounds

Fig. 5: Paths chosen in network topology-1

We consider two network topologies, both of which involve three sensors and one source. The sensing parameters $\boldsymbol{\beta}$ are chosen as inversely proportional with the squared distances between the sensor and the source, *i.e.,* for source and $m^{th}$ sensor locations $\boldsymbol{x}_s$ and $\boldsymbol{x}_m$, we have $\beta_m = b \times ||\boldsymbol{x}_s - \boldsymbol{x}_m||_2^{-2}$ for some $b \in \mathbb{R}^+$. The channel parameters, $\boldsymbol{\alpha}$, are determined similarly: given the AV location $\boldsymbol{x}_a$, we have $\alpha_m = a \times ||\boldsymbol{x}_a - \boldsymbol{x}_m||_2^{-2}$, for a given $a \in \mathbb{R}^+$. The path step size is set to 0.01 and paths are of length 30 steps, and the weight vector $r = [1, 1, 1]$. The AV is set to point $[-1, 0]$ on a $3 \times 3.5$ grid and source location is set to $[1.5, 0]$. We provide two examples that demonstrate different aspects of path selection and metrics.

We first consider a network topology similar to the motivating example in Fig. 1, where an AV gathers data from three deployed sensors. We choose $a = b = 10$, and $P_m = 10$ for each sensor, hence $P_T = 30$, and $\gamma = [1, 1, 4]$ to capture the effect of non-symmetry in the network topology. We plot the paths chosen by upper bounds metrics, in Fig. 5(a) and ones with the lower bounds, in Fig. 5(b) Here, the upper bounds (obtained via the AF scheme) of the SR metric yields a path towards the sensors closest to the source, as shown in Fig. 5(a). Power optimization makes this path only more skewed toward the closest sensor (sensor-1), which is theoretically expected since sensor 1 *senses* the source with minimum distortion, and hence more power is allowed for sensor-1 in the optimized power allocation method. The upper bound of the FR metric (achieved via the AF scheme) results in a path towards sensor-3, which is due to fact that sensor-3 represents a larger area than other sensors and, hence the asymmetry in $\gamma$. Therefore, the numerical path selection results demonstrated in Fig. 5(a) confirm our intuition in the example in Fig. 1. However, for the lower bounds, all four metrics yield the same path towards to the closest sensor to the AV (sensor-2). *The results in this example indicate a very interesting conclusion on the importance of the metric selection, since the optimal paths in this example strongly depend on the metric chosen.*

Next, we change the network topology to a semi-symmetric version where two of the sensors (sensors 1 and 3) are equally distant from the source. In this example, we analyze the impact of channel parameters on the metrics and eventually path selection. Since the settings is relatively more symmetric (as opposed to previous setting), we set $\gamma = [1, 1, 1]$ and $r = [1, 1, 1]$. We plot the paths found for parameter values $a = 10, b = 1$, and $P_m = 100$ for all $m$ in Fig. 6. Note that this example presents an interesting case for path selection, since paths chosen by different metrics vary widely. First, let us explain why $J_S^U(\boldsymbol{P})$ and $J_F^U(\boldsymbol{P})$ yield these paths. An obvious question is the following: why do these paths tend to move away from all the sensors and source in the beginning of the path? The answer lies in the network parameters and topology. This example involves a very small sensing parameter, ($b$), compared to communication parameters, ($a$ and $P$), which implies that the sensor with the worst sensing channel can even amplify overall noise at the AV, *i.e.,* sensor-2 output interferes with the source $S$, in SR case, or with other sensor observations $U_m$ in the FR case. Note that AV can only change its communication channel quality while the sensing channel is fixed in the problem setting. This indicates that the AV can only mitigate the effect of these "bad sensors" (here sensor-2) by moving away from them. This is exactly what we observe in Fig. 6(a) for the paths generated by $J_S^U(\boldsymbol{P})$ and $J_F^U(\boldsymbol{P})$. Note that due to symmetry overall costs (when measured by the same metric) of both paths are identical (in Fig. 6(a), these two paths with identical costs are assigned to $J_S^U(\boldsymbol{P})$ and $J_F^U(\boldsymbol{P})$ randomly). When power is optimized, sensor-2 is not allowed to decrease the communication channel SNR at the AV ($P_2 \ll 10$), hence the paths by $J_S^U(P_T)$ and $J_F^U(P_T)$ are towards to the middle of all sensors due to symmetry. Note that all lower bounds yields the same path for this example.

Next, in the same network topology, we increase the sensing parameter $b$ and decrease the communication parameters, $a$ and $P$, specifically, we set $a = 1, b = 10, P_m = 10$ for all $m$, and hence $P_T = 30$, and keep other parameters the same as the previous setting ($r = [1, 1, 1]$, $\gamma = [1, 1, 1]$). For these settings, as Fig. 7 demonstrates, all metrics yield a path toward the closest sensor to AV, which is sensor-2 in this setting. This is theoretically expected since the bottleneck for the performance here is the communication over MAC, as opposed to the sensing channel (which was the case in the previous setting in Fig. 6) due to channel parameters.

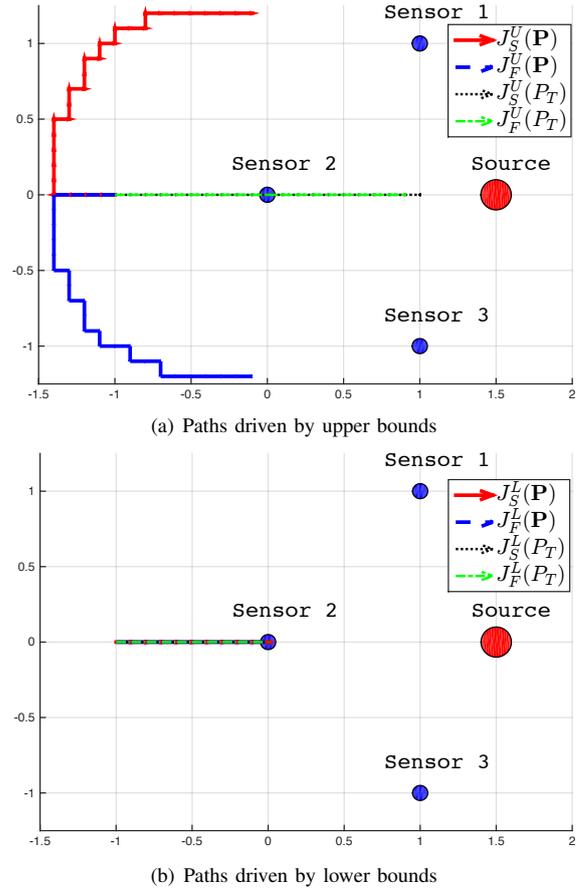

(a) Paths driven by upper bounds

(b) Paths driven by lower bounds

Fig. 6: Paths chosen in network topology-2, small sensing parameters

## V. Conclusions

In this paper, we have analyzed bounds of communication performance over Gaussian sensor networks for path planning problems. We have considered two main metrics: i) the underlying source is estimated, and ii) the spatial field is reconstructed. We have provided a unified proof for the upper and lower bounds of the fundamental limits of communication with these metrics, for fixed and optimized power allocations. Finally, the effect of metric selection, network topology and channel parameters on the selected path is analyzed. Our results show that depending on the network, metric selection and power optimization may significantly impact the optimal path in data gathering. Simulation results suggest that the metrics associated with the lower bounds seem to be more sensitive to channel parameters and network topology than those of the outer bounds.

This paper is concerned static sensors and a mobile data gathering device, and a single scalar source and scalar channels. Our future work includes extension to dynamic and multi-source and multi-dimensional settings, and on the optimal resource (power) allocation in time over a fixed path period, and its impact on path selection. Analysis of the settings where sensing is performed in a mobile platform (see *e.g.,* [33]), or partially known or timely varying network statistics are left as future work. We finally note that this paper highlights the need for further information-theoretic analysis of fundamental bounds for sensor networks. One research direction is to utilize structured codes [34] in such networked source-channel coding problems.

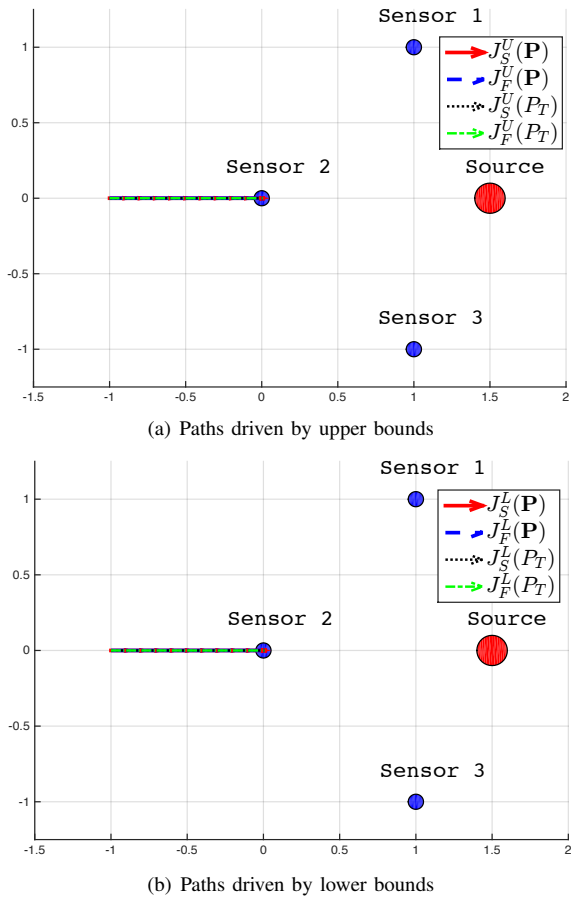

(a) Paths driven by upper bounds

(b) Paths driven by lower bounds

Fig. 7: Paths chosen in network topology-2, small communication parameters

## APPENDIX A
## GAUSSIAN REMOTE COMPRESSION PROBLEM

In this problem, an underlying Gaussian source $S \sim \mathcal{N}(0, 1)$ is observed under additive noise $\boldsymbol{W} \sim \mathcal{N}(\boldsymbol{0}, R_W)$ as $\boldsymbol{U} = S + \boldsymbol{W}$. These noisy observations, *i.e.*, $\boldsymbol{U}$, must be encoded in such a way that the decoder produces a good approximation to the original underlying source. This problem was proposed in [35] and solved in [36] (see also [37]). A lower bound for this function for the non-Gaussian sources within the symmetric setting where all $U$'s have identical statistics was presented in [38]. Here, we simply extend the results in [36] to our asymmetric setting, noting

$$D = \mathbb{E}\{(S - \hat{S})^2\}, \tag{65}$$

$$R = \min I(\boldsymbol{U}; \hat{S}), \tag{66}$$

where $\boldsymbol{U} = \boldsymbol{\beta}S + \boldsymbol{W}$, $\boldsymbol{W} \sim \mathcal{N}(\boldsymbol{0}, R_W)$, and $R_W$ is an $M \times M$ identity matrix. The minimization in (66) is over all conditional densities $p(\hat{s}|\boldsymbol{u})$ that satisfy (65). The MSE distortion can be written as sum of two terms

$$D = \mathbb{E}\{(S - T + T - \hat{S})^2\}, \tag{67}$$

$$= \mathbb{E}\{(S - T)^2\} + \mathbb{E}\{(T - \hat{S})^2\}, \tag{68}$$

where $T \triangleq \mathbb{E}\{S|\boldsymbol{U}\}$. Note that (68) holds since

$$\mathbb{E}\{(S - T)(\hat{S} - T)\} = 0, \tag{69}$$

as the estimation error, $S - T$, is orthogonal to any function[3] of the observation, $\boldsymbol{U}$. The estimation distortion

$$D_{est} \triangleq \mathbb{E}\{(S - T)^2\} \tag{70}$$

is constant with respect to $p(\hat{s}|\boldsymbol{u})$. Hence, the minimization is over the densities that satisfy a distortion constraint of the form $\mathbb{E}\{(T - \hat{S})^2\} \leq D_{rd}$ and $R = \min I(\boldsymbol{U}; \hat{S})$. Hence, we write (68) as

$$D = D_{rd} + D_{est}. \tag{71}$$

Note that due to their Gaussianity, $T$ is a sufficient statistic of $\boldsymbol{U}$ for $S$, *i.e.*, $S - T - \boldsymbol{U}$ forms a Markov chain in that order and $T \sim \mathcal{N}(0, \sigma_T^2)$. Hence, $R = \min I(\boldsymbol{U}; \hat{S}) = \min I(T; \hat{S})$ where minimization is over $p(\hat{s}|t)$ that satisfy $\mathbb{E}\{(T - \hat{S})^2\} \leq D_{rd}$, where all variables are Gaussian. This is the classical Gaussian rate-distortion problem, and hence:

$$D_{rd}(R) = \sigma_T^2 2^{-2R}. \tag{72}$$

Note that $T = R_{SU}R_U^{-1}\boldsymbol{U}$, where $R_{SU} \triangleq \mathbb{E}\{S\boldsymbol{U}^T\}$ and $R_U$ is given in (37). Note that $R_U$ is structured, and can easily be manipulated. We compute $\sigma_T^2$ as

$$\sigma_T^2 = R_{SU}R_U^{-1}R_{SU}^T = \frac{\sum_{m=1}^{M} \beta_m^2}{1 + \sum_{m=1}^{M} \beta_m^2}, \tag{73}$$

and using standard linear estimation principles, we obtain

$$D_{est} = \frac{1}{1 + \sum_{m=1}^{M} \beta_m^2}. \tag{74}$$

Plugging (74) in (72) and using (71) yields

$$D = \left( \frac{1}{1 + \sum_{m=1}^{M} \beta_m^2} + \frac{\sum_{m=1}^{M} \beta_m^2}{1 + \sum_{m=1}^{M} \beta_m^2} 2^{-2R} \right). \tag{75}$$

## APPENDIX B
## GAUSSIAN VECTOR SOURCE CODING

The problem of interest to find $D(R)$ that minimize $D = \sum_{m=1}^{M} \gamma_m \mathbb{E}\{(U_m - \hat{U}_m)^2\}$, over $R_m \geq 0$ subject to $R = \sum_{m=1}^{M} R_m$. This is a variant of a standard problem of encoding multiple independent Gaussian variables [39]. Note that $R_U$ given in (37) accepts the eigen-decomposition $R_U = Q_U^T \Lambda Q_U$, where $\Lambda$ is a diagonal matrix with entries $1, 1, \ldots, 1, 1 + \sum_{m=1}^{M} \beta_m^2$ and $Q_U$ is a unitary matrix. Hence, the problem can be converted (by linear transformation) to that of encoding independent Gaussian scalars with variances $1, 1, \ldots, 1 + \sum_{m=1}^{M} \beta_m^2$, i.e., we minimize $D = \sum_{m=1}^{M} \gamma_m' D_m$ subject to $R = \sum_{m=1}^{M} R_m$, where

$$\gamma_k' \triangleq [Q_U^T \operatorname{diag}(\boldsymbol{\gamma}) Q_U]_{kk} \tag{76}$$

and $R_m = \left( \frac{1}{2} \log \left( \frac{\Lambda_m}{D_m} \right) \right)^+$. Equivalently, we minimize

$$J = \sum_{m=1}^{M} \frac{1}{2} \log \left( \frac{\Lambda_m}{D_m} \right) + \lambda \sum_{m=1}^{M} \gamma_m' D_m \tag{77}$$

over the set of $D_m$ that satisfy $D_m \leq \Lambda_m$ (hence, $R_m \geq 0, \forall m$). Applying the KKT conditions, we have

$$\frac{\partial J}{\partial D_m} = -\frac{1}{2} \frac{1}{D_m} + \lambda \gamma_m' = 0, \tag{78}$$

or

$$D_m = \frac{1}{2 \lambda \gamma_m'} \triangleq \theta / \gamma_m'. \tag{79}$$

---

[3]Note that $\hat{S}$ is also a deterministic function of $\boldsymbol{U}$, since the optimal reconstruction can always be achieved by deterministic codes.

Hence, we have

$$R = \frac{1}{2} \sum_{m=1}^{M} \log \left( \frac{\Lambda_m}{D_m} \right), \tag{80}$$

where

$$D_m = \min\{\theta / \gamma_m', \Lambda_m\}, \tag{81}$$

and $\theta$ is chosen so that $D = \sum_{m=1}^{M} \gamma_m' D_m$, for $\operatorname{diag}(\boldsymbol{\gamma}') = Q_U^T \operatorname{diag}(\boldsymbol{\gamma}) Q_U$. The *water-filling* nature of the solution (see (81)) prevents the achievement of closed-form solutions for the power-distortion curve. To provide insight, we assume the "high rate" (low distortion) regime, where each component is effective, i.e., $\theta \leq \min_m \{\Lambda_m \gamma_m'\}$. With this assumption, we have $D_m = \theta / \gamma_m'$, and hence

$$\theta = D / M. \tag{82}$$

Plugging (82) in (80), we have

$$R = \frac{1}{2} \sum_{m=1}^{M} \log \left( \Lambda_m \gamma_m' \right) - \frac{M}{2} \log \left( \frac{D}{M} \right), \tag{83}$$

and noting that $\sum_{m=1}^{M} \log \left( \Lambda_m \gamma_m' \right) = \log \left( \prod_{m=1}^{M} \Lambda_m \gamma_m' \right) = \log \left( (1 + \sum_{m=1}^{M} \beta_m^2) \prod_{m=1}^{M} \gamma_m' \right)$, we have

$$D = M \left( (1 + \sum_{m=1}^{M} \beta_m^2) \prod_{m=1}^{M} \gamma_m' \right)^{\frac{1}{M}} \exp \left\{ -\frac{2}{M} R \right\}. \tag{84}$$

## Appendix C
## Witsenhausen's Lemma

**Lemma 1** (from [28]). *Consider two sequences of i.i.d. random variables $X(i)$ and $Y(i)$, generated from a joint density $P_{X,Y}$, and two arbitrary functions $f, g : \mathbb{R} \to \mathbb{R}$ satisfying*

$$\mathbb{E}\{f(X)\} = \mathbb{E}\{g(Y)\} = 0, \quad \mathbb{E}\{f^2(X)\} = \mathbb{E}\{g^2(Y)\} = 1. \tag{85}$$

*For any functions $f_N, g_N : \mathbb{R}^N \to \mathbb{R}$ satisfying*

$$\mathbb{E}\{f_N(\boldsymbol{X})\} = \mathbb{E}\{g_N(\boldsymbol{Y})\} = 0, \quad \mathbb{E}\{f_N^2(\boldsymbol{X})\} = \mathbb{E}\{g_N^2(\boldsymbol{Y})\} = 1, \tag{86}$$

*for length $N$ vectors $\boldsymbol{X}$ and $\boldsymbol{Y}$, we have*

$$\sup_{f_N, g_N} \mathbb{E}\{f_N(\boldsymbol{X}) g_N(\boldsymbol{Y})\} \leq \sup_{f, g} \mathbb{E}\{f(X) g(Y)\}. \tag{87}$$

*Moreover, the supremum above is attained by linear mappings, if $P_{X,Y}$ is Gaussian density.*